\documentstyle[12pt,aasms4]{article}

\received{December 30 1996}
\accepted{April 20 2000}

\begin{document}

\title{Environmental Effect on the Associations of Background
       Quasars with Foreground Objects: II. Numerical Simulations}

\author{Xiang-Ping Wu and Xiao-Hong Zhu}
\affil{Beijing Astronomical Observatory, Chinese Academy of Sciences,
       Beijing 100080, China}

\author{Yi-Peng Jing}
\affil{Research Center for the Early Universe,
School of Science, University of Tokyo, Bunkyo-ku, Tokyo 113, Japan}

\and

\author{Li-Zhi Fang}
\affil{Department of Physics, University of Arizona, Tucson, AZ 85721}

\begin{abstract}
Using numerical simulations of cluster formation in the standard CDM
model (SCDM) and in a low-density, flat CDM model with a cosmological
constant (LCDM), we investigate the gravitational lensing explanation
for the reported associations between background quasars and
foreground clusters. Under the thin-lens approximation and the
unaffected background hypothesis , we show that the recently
detected quasar overdensity around clusters of galaxies on scales of
$\sim10$ arcminutes cannot be interpreted as a result of the
gravitational lensing by cluster matter and/or by their environmental
and projected matter along the line of sight, which is consistent with
the analytical result based on the observed cluster and galaxy
correlations (Wu, et al. 1996). It appears very unlikely that
uncertainties in the modeling of the gravitational lensing can account for
the disagreement between the theoretical predictions and the
observations. We conclude that either the detected signal of the
quasar-cluster associations is a statistical fluke or the associations
are generated by mechanisms other than the magnification bias.
\end{abstract}

\bigskip


\section{Introduction}

Recently, a statistically significant correlation between distant
quasars and nearby clusters of galaxies is detected on scales of
$\sim10$ arcminutes (Rodrigues-Williams \& Hogan, 1994; Wu \& Han
1995; Rodrigues-Williams \& Hawkins, 1995; Seitz \& Schneider 1995).
It seems very unlikely that this correlation is due to
the gravitational lensing of the quasars by the clusters, unless a
considerably large velocity dispersion of $\sigma_v>2000$ km s$^{-1}$
is assumed for the clusters.  This is because the association scales of
$\sim10^{\prime}$ at $z\sim0.2$ correspond to the ``edges'' of
clusters of galaxies, where the influence of the gravitational lensing
by cluster matter alone becomes small. Motivated by the remarkable
quasar number excess around clusters, an attempt has been made to
attribute the quasar-cluster associations to the gravitational lensing
by the large-scale structures traced by clusters of galaxies, namely,
the cluster environmental effect (Wu \& Fang 1996a; Wu et al. 1996;
hereafter Paper I and Paper II). A similar scenario has ever been
suggested for the quasar-galaxy angular correlations on large-scales
(Bartelmann \& Schneider 1993).  It appears that in the framework of
gravitational lensing the cluster environmental matter described by
the cluster-cluster and cluster-galaxy two-point correlation functions
is insufficient to account for the quasar overdensity around clusters,
if one adopts the unaffected background hypothesis, i.e., the observed
quasar number counts as a whole have not been seriously contaminated
by gravitational lensing.

However, a number of issues regarding the matter clustering around
galaxy clusters may be overlooked if we only employ the two-point
correlation functions.  First, the biasing of the luminous matter with
respect to the dark matter is not concerned. Second, a singular
isothermal sphere model was presumed for the matter distribution of
clusters, which is oversimple because of the presence of
substructures.  Third, the two-point cluster-cluster correlation function 
is inappropriate for the description of matter clustering within a
distance of $\sim5$ Mpc from clusters. Yet, these problems can be
easily dealt with by means of the cosmological numerical
simulations. Indeed, clusters of galaxies usually reside in the
intersections of filaments and pancakes, and the simulations of
cluster formation provide an effective way to probe the environmental
matter distributions traced by clusters. Alternatively, numerical
simulations allow us to map all the matter inhomogeneities along the
line of sight to the distant sources, giving rise to an estimate of
the amplitude of the gravitational lensing effect by these matter
clumps, i.e., the projection effects. It has been shown in a recent
numerical study (Cen 1996) that the projection effects may
significantly contaminate a number of physical quantities of clusters.
Therefore, simulations can largely complement to analytic
investigations in the study of cluster properties.  In this paper we
study the gravitational lensing effect on the quasar-cluster
associations using a set of cosmological numerical simulations.
Similar numerical techniques have been employed in the study of the
gravitational lensing by microlenses (see, for example, Schneider et
al. 1992), by clusters of galaxies (e.g. Bartelmann \& Schneider 1991)
and by large-scale structures (e.g. Cen et al. 1994; Wambsganss et al.
1996).

As the first step towards investigating the environmental effects on
the quasar-cluster associations with numerical simulations, we adopt the
``thin'' lens approximation, i.e., we project all the matter
inhomogeneities along the line of sight onto the lens (cluster) plane
and make no distinction between the environmental effects and the
projection effects. A more sophisticated treatment of the gravitational
lensing effect by clusters and large-scale structures is to use the
approximation of the multiple lens planes and the ray-shooting
technique (Wambsganss et al. 1996; reference therein).  For a
transparent object at cosmological distance, its lensing
magnification becomes significant only if its surface mass density
$\Sigma$ is comparable to the critical value of $\Sigma_{crit}=(c^2/4\pi
G)(D_s/D_dD_{ds})$ (Turner et al. 1984), where $D_d$, $D_s$ and
$D_{ds}$ are the angular diameter distances to the lens (cluster), to
the background source (quasar) and from the lens to the source,
respectively. To account for the reported quasar overdensity around
clusters on scales of $\sim10^{\prime}$ in terms of the gravitational
lensing, it has been shown that a surface mass density of
$\Sigma^*\approx0.2\Sigma_{crit}$ is required (Paper I). We first
examine whether the projected cluster environmental matter can reach a
value as high as $\Sigma^*$ (section 2).  We then present a
detailed computation of the magnification patterns induced by all the
matter clumps and compute their resulting quasar enhancement factor
$q$ (section 3).  Finally, we briefly discuss and summarize our
results (section 4). Throughout the paper, we adopt a Hubble constant
of $H_0=100$ km s$^{-1}$ Mpc$^{-1}$ and a flat cosmological model of
$\Omega_0+\lambda_0=1$, where $\Omega_0$ and $\lambda_0$ denote 
the density parameters contributed by the cold dark matter and by the
cosmological constant, respectively.

\section{Cluster environmental matter distributions}

We work with two cosmological models: the standard CDM model of
$\Omega_0=1$ and $\lambda_0=0$ (SCDM) and a low-density, flat CDM
model with a nonzero cosmological constant of $\lambda_0=0.7$ 
($\Omega_0=0.3$) (LCDM).  Their primordial density fluctuations are
normalized by $\sigma_8=0.6$ for SCDM and $\sigma_8=1$ for LCDM, where
$\sigma_8$ is the present rms mass contrast of a sphere of radius
$8$ Mpc.  The normalizations are chosen such that both models predict
cluster abundances similar to those observed (Jing \& Fang 1994).
Although SCDM is inconsistent with many observed properties of
clusters of galaxies (e.g. Bahcall \& Cen 1993), the model is still
widely adopted as a working theory since it is simple and can
qualitatively give us a sense of how the cosmic structures may look
like. On the other hand, LCDM is one of the prevailing models that can
quantitatively fit nearly all the observed properties of clusters
including the mass function, the velocity dispersion distribution, the
two-point correlation function and substructures around clusters
(e.g. Bahcall \& Cen 1993; White et al. 1993; Jing \& Fang 1994; Jing
et al. 1995; Kitayama \& Suto 1996; Boute \& Xu 1997). 

We use a P$^3$M N-body code to generate the numerical simulations.  For
a detailed description of the simulations and of our identification of
clusters, the reader is referred to Jing \& Fang (1994).  The simulations are
performed in a cubic box of $128^3$ Mpc$^3$ with periodic
boundaries. A total of $64^3$ particles is utilized and each particle
has mass of $2.2\times10^{12}\Omega_0$ M$_{\odot}$. The force
resolution in our simulations is $\sim0.1$ Mpc, and each rich cluster
like the Abell one is composed of more than 100 particles. Since what we
are interested in is the large angular correlations between quasars
and clusters rather than the arclike images of background galaxies,
high resolutions are not necessarily needed. To achieve a good
statistical significance, we have run three realizations for SCDM and
five realizations for LCDM, which yield roughly the same cluster
populations.

For the present work, the most important quantity is the mass density
$\rho(r)$ and the projected surface mass density $\Sigma(r)$. They are
obtained using the Gaussian smoothing kernel $W(r,s)$ with a smoothing
length $s$ equal to the local mean particle separation. As examples,
Fig.1 and Fig.2 display the 2-D mass distributions around a cluster of
galaxies in SCDM and in LCDM respectively, produced by projecting a
rectangular cylinder of 60 Mpc $\times$ 60 Mpc $\times$ 75 Mpc, where
two surface mass density ``filters'' are employed:
(a)$\Sigma\geq\Sigma^*$ and (b)$\Sigma\geq0.1\Sigma^*$.  The thickness
of 75 Mpc is adopted to match the maximum separation in the
cluster-cluster correlation function (Postman et al.  1992). It is
immediately apparent from Fig.1 and Fig.2 that very sparse areas
around clusters meet the requirement of $\Sigma\geq\Sigma^*$, i.e.,
clusters of galaxies do not inhabit the dense matter environments. In
practice, all the points shown in Fig.1(a) and Fig.2(a) correspond to the
cores of groups and clusters.  The overall mean 2-D mass density in
each field turns to be $\sim10^{-3}$ g cm$^{-2}$, in accordance with
our analytic estimate of the mean cluster environmental mass density
(Paper II). So, our first intuitive impression based on the 2-D mass
distributions in the vicinity of clusters is that there exists no
massive uniform sheet around each cluster.  However, it should be
noted that this does not exclude the possibility of attributing the
quasar-cluster associations to the result of gravitational
lensing. Indeed, the naive lensing model of a uniform mass sheet for
cluster environmental matter distribution needs to be improved. There
are numerous systems such as groups and poor clusters in the cluster
fields [see Fig.1(b) and Fig.2(b)] and we have to investigate whether
their combined lensing magnifications are capable of producing the
observed quasar overdensity behind clusters.

\placefigure{fig1}
\placefigure{fig2}

\section{Magnification patterns and quasar overdensity}

We now determine the deflection angle of light $\mbox{\boldmath $\alpha$}$
at a position $\mbox{\boldmath $\theta$}=(\theta_x,\theta_y)$
from the cluster center  $\mbox{\boldmath $\theta$}=0$ by all the matter 
projected onto the cluster plane.  To do this, we treat the matter 
distribution inside a cell as a uniform mass sheet 
which takes the value of the surface mass density at the grid.
The total $\mbox{\boldmath $\alpha$}$ can be obtained by 
\begin{equation}
\mbox{\boldmath $\alpha$}=\frac{4G}{c^2}D_d\displaystyle\sum_i
      \sigma_i\int_{s_i}\frac{\mbox{\boldmath $\theta$}-
                              \mbox{\boldmath $\theta$}_i}
                             {|\mbox{\boldmath $\theta$}-
                               \mbox{\boldmath $\theta$}_i|^2}
                        d^2\mbox{\boldmath $\theta$}_i,
\end{equation}
where the integration and summation are performed inside each cell $s_i$
with surface mass density $\sigma_i$  
and over all the cells on the cluster plane, respectively.
The lens equation for a background source at redshift $z_s$ and with
angular position  $\mbox{\boldmath $\beta$}$  is simply
\begin{equation}
\mbox{\boldmath $\beta$}-\mbox{\boldmath $\theta$}
=\frac{D_{ds}}{D_s}\mbox{\boldmath $\alpha$}.
\end{equation}
The Jacobian 
$\partial\mbox{\boldmath $\beta$}/\partial\mbox{\boldmath $\theta$}$
yields the magnification of an image at $\mbox{\boldmath $\theta$}$
of the background source 
\begin{equation}
\mu(\mbox{\boldmath $\theta$})=
\left[1-\frac{\phi^2+\psi^2}{(\pi \Sigma_{crit})^2}\right]^{-1},
\end{equation}
where
\begin{equation}
\phi\equiv \displaystyle\sum_i\sigma_i\int_{s_i}
                       \frac{(\theta_x-\theta_{ix})^2-
                             (\theta_y-\theta_{iy})^2}
                             {|\mbox{\boldmath $\theta$}-
                               \mbox{\boldmath $\theta$}_i|^4}
                        d^2\mbox{\boldmath $\theta$}_i,
\end{equation}
and
\begin{equation}
\psi\equiv -\displaystyle\sum_i\sigma_i\int_{s_i}
                       \frac{2(\theta_x-\theta_{ix})
                              (\theta_y-\theta_{iy})}
                             {|\mbox{\boldmath $\theta$}-
                               \mbox{\boldmath $\theta$}_i|^4}
                        d^2\mbox{\boldmath $\theta$}_i.
\end{equation}

For a given position $\mbox{\boldmath $\theta$}$, we calculate 
the magnification due to all the matter distributed within a square
region of 15 Mpc $\times$ 15 Mpc surrounding 
$\mbox{\boldmath $\theta$}$. That is, a total of $150\times150$
cells has been used. Two examples of the matter distributions 
and the corresponding magnification patterns in a field of
$15$ Mpc $\times15$ Mpc centered at a cluster are shown
in Fig.3 and Fig.4 for SCDM and LCDM, respectively. It is evident
that, although all the matter in the field contributes to the
magnification at a given position, the magnification patterns
essentially follow the matter distributions. The high magnification
usually appears in the cores of clusters, giving rise to 
the strong lensing events such as the arclike images of background 
galaxies.

\placefigure{fig3}
\placefigure{fig4}

Knowing the magnification patterns around each cluster, 
we are able to statistically compute the amplitude of 
quasar overdensity behind an ensemble of clusters due to 
the magnification bias. 
The mean quasar enhancement factor $q$ for a quasar limiting magnitude $B$
and within a projected distance $r$ from a cluster center is
\begin{equation}
q=\frac{\int_o^r q_{{\rm local}}[B,\mu(r)]2\pi rdr}{\pi r^2},
\end{equation}
in which $q_{{\rm local}}$ represents the local quasar enhancement factor 
(Narayan 1989):
\begin{equation}
q_{{\rm local}}=\frac{N_q(<B+2.5\log\mu)}{N_q(<B)}\frac{1}{\mu},
\end{equation}
and $N_q(<B)$ is the quasar number-magnitude relation. Supposing that
the observed quasar number counts as a whole are unaffected by 
gravitational lensing (i.e. the unaffected background hypothesis), 
we can utilize the Boyle et al. (1988) quasar
counts to estimate the value of $q$. Here, the radio-selected
quasars and the variability-selected ones (Hawkins \& V\'eron 1993)
are not included.

Taking the typical redshifts of $z_s=2$ for the background quasars and
$z_d=0.2$ for the foreground clusters and using a limiting magnitude
of $B<18.5$ which is comparable to the one in  
the measurements of quasar-cluster associations,
we have calculated the enhancement factors around 30 rich clusters of galaxies
selected randomly from our cluster catalogs with cluster masses ranging
from $7.0\times10^{14}M_{\odot}$ to  $1.5\times10^{15}M_{\odot}$ for
SCDM and from $3.1\times10^{14}M_{\odot}$ to  
$1.2\times10^{15}M_{\odot}$ for LCDM, respectively.
The mean value of $q$ as a function of the projected distance
from cluster centers has been illustrated in Fig.5. Aside from 
a weakly positive correlation between clusters and quasars at the
central regions of clusters, we have not detected a remarkable 
overdensity of quasars around clusters out to cluster radii.
This conclusion holds true for both SCDM and LCDM models. 
It appears unlikely that the statistical fluctuations in our simulations 
can account for the large discrepancy between the theoretically expected 
enhancements  $q\approx1$ and the observationally reported values 
$q\approx2$ at a comoving distance of $r\approx1$ -- 10 Mpc. 

\placefigure{fig5}

\section{Discussion and conclusions}

The present numerical study of the gravitational lensing effects by
clusters and their environmental matter, together with our previous
analytic investigations (Paper I and II), has resulted in a weak
correlation between background quasars and foreground clusters of
galaxies on scales of $\sim10^{\prime}$. This suggests that the
recently reported overdensity of quasars around clusters
out to several cluster radii is not the result of the gravitational
magnification bias unless (1) clusters of galaxies have a relatively
large velocity dispersion up to $5000$ km s$^{-1}$ or (2) the
observed quasar number-magnitude relation has been seriously
contaminated by gravitational lensing.  Recall that the similar
conditions were required in order to account for the quasar-galaxy
associations even on small scales of a few arcseconds (e.g. Webster et
al. 1988; Narayan 1989).  The first possibility implies that the
cluster masses required for the lensing explanation of the quasar-cluster
associations are of an order of magnitude higher than the known
dynamical cluster masses. Although the dynamical analyses based upon
hydrostatic equilibrium may underestimate cluster masses by a factor
of $\sim2$ as compared to the gravitational lensing method using
arcs/arclets and weak lensing phenomena (Wu \& Fang 1996b; reference
therein), it is very unlikely that this mass discrepancy can be as
large as 10 ! Therefore, such a possibility can be definitely
excluded.  As for the second possibility, the previous work (Schneider
1992; Pei 1995) has shown that the contamination of the quasar number
counts from gravitational lensing by galactic matter is trivial.
However, in the case of the quasar-cluster associations, the
association area is usually a few tens percent of the total searching
field, i.e., the association quasars are very common. So, the question
remains open whether it is reasonable to inherit the unaffected
background hypothesis in the study of the quasar-cluster associations.

Our estimate of the quasar enhancement factor may suffer from 
a number of uncertainties.  The thin-lens approximation is in principle 
inappropriate for the description of large-scale matter distributions.
It will be necessary in our subsequent studies to adopt a more realistic
model that is composed of multiple lens planes. Also, we have used a 
uniformly smoothed mass distribution inside a cell in the calculation of
deflection angle instead of the usually adopted pointlike model.
In the latter case, all the mass inside a cell is assigned to its center.
Our treatment avoids the occurrence of the artificially-induced strong 
magnifications near the center of each cell (point mass). 
However,  numerical simulations with high resolutions
will be needed  to overcome the softening of lensing 
ability due to the relatively large size (0.1 Mpc $\times$ 0.1 Mpc) 
for a cell.

\null From the observational point of view, 
although each measurement of the quasar-cluster associations claims that
the detected quasar overdensity around clusters is not the result of 
statistical fluctuations or does not suffer from  other observational 
selection effects,  the real situations may be complicated.
Among the four measurements, three different approaches are used 
to describe quantitatively the quasar-cluster associations because of 
the difficulty of obtaining the undisturbed or background quasar 
surface number density. If the definition of Seitz \& Schneider (1995)
is adopted, the quasar enhancement factors  found by 
Rodrigues-Williams \& Hogan (1994) and Rodrigues-Williams \&
Hawkins (1995) should be somewhat reduced. This may partially remove
the discrepancy between the theoretical expectations and the observations. 
Alternatively, the null/negative associations were also detected in
some searches (Wu \& Han 1995; Seitz \& Schneider 1995). However, 
those results were attributed to other mechanisms and hence, have not 
been included in the lensing analysis. It appears that
current measurements of the quasar-cluster associations
are probably biased. Actually,     
in contrast to the positive result, an anti-correlation between
high redshift quasars and foreground clusters was reported 
many years ago (Boyle et al. 1988), which was interpreted as the 
result of the obscuration 
by the intracluster dust. As a number of subsequent 
observations has also provided the evidence against the excess number of 
quasars in the vicinity of clusters, 
further observations will be needed to improve the 
confidence level of different results before any definite conclusions can be
drawn. Considering the same status and  difficulty for
the measurements and explanations of the quasar-galaxy associations 
(Zhu et al. 1997; Fried 1997; reference therein)
and the quasar-quasar associations (Burbidge et al. 1997), 
we believe that all the association problems reported thus far 
might have a common origin: either the observations
have not detected the associations of background quasars
with foreground objects as a result of gravitational lensing, 
or the associations are generated by 
some mechanism other than the gravitational lensing if the observed 
quasar number counts are not very much different from the intrinsic ones.  

Finally, because the discrepancy in the quasar
enhancement factors between the theoretical
expectations in terms of gravitational lensing and the observations 
is very large,  it is unlikely that the conclusions
reached in this paper can be  significantly affected by 
utilizing other cosmological models instead of SCDM and LCDM

\acknowledgments

We are grateful to an anonymous referee for helpful criticisms.   
Y.P.J acknowledges the receipt of a JSPS Postdoctoral Fellowship.
This work was supported by the National Science Foundation
of China and by Monbusho Grant-in-aid for JSPS Fellows No.XXXX.

\clearpage

\figcaption{An example of the SCDM generated
2-D matter distributions in a square region of
60 Mpc $\times$ 60 Mpc centered at a cluster, obtained by projecting
all the matter inside a cube with comoving length of $75$ Mpc. 
(a)Surface mass density $\Sigma\geq\Sigma^*$ and (b) $\Sigma\geq0.1\Sigma^*$,
where $\Sigma^*$ is the surface mass density required to
produce the reported amplitude of quasar overdensity around clusters
of galaxies in the framework of gravitational lensing.
             \label{fig1}}

\figcaption{The same as Fig.1 but for LCDM.
             \label{fig2}}

\figcaption{An example of the 2-D matter distribution (a) 
and the corresponding magnification patterns 
(b) around the same cluster shown in Fig.1(b) 
in a square region of 15 Mpc $\times$ 15 Mpc. The void regions,  
the shadows and the darkened areas in (b) 
represent $\mu<1.001$, $1.001\leq\mu\leq1.005$ and
$\mu>1.005$, respectively.
 \label{fig3}}

\figcaption{The same as Fig.3 but for the cluster in Fig.2.
             \label{fig4}}

\figcaption{The enhancement factor $q$ of optically selected quasars 
with $B\leq18.5$ versus searching distance $r$ from cluster centers.
A total of 30 rich clusters has been used in each model (SCDM and LCDM) and
$1\sigma$ error bars have been shown.
             \label{fig5}}


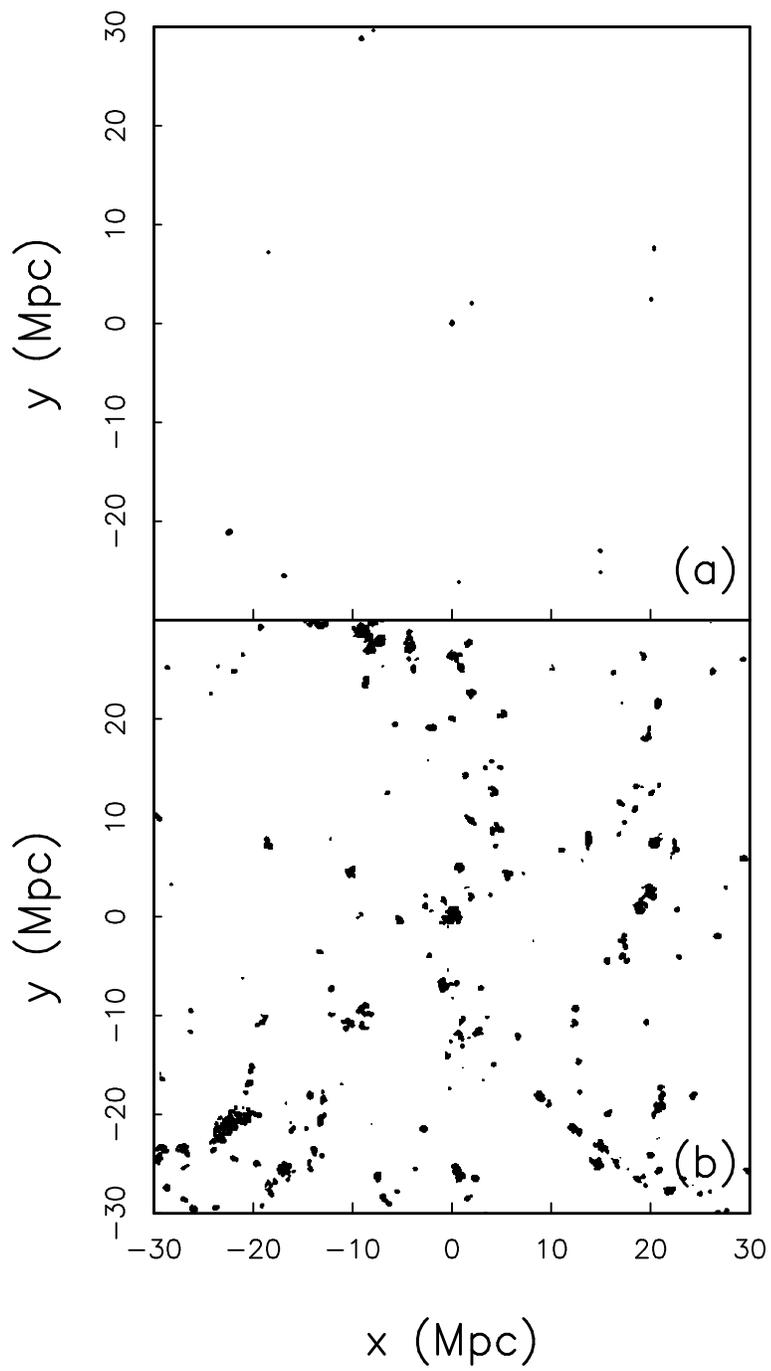
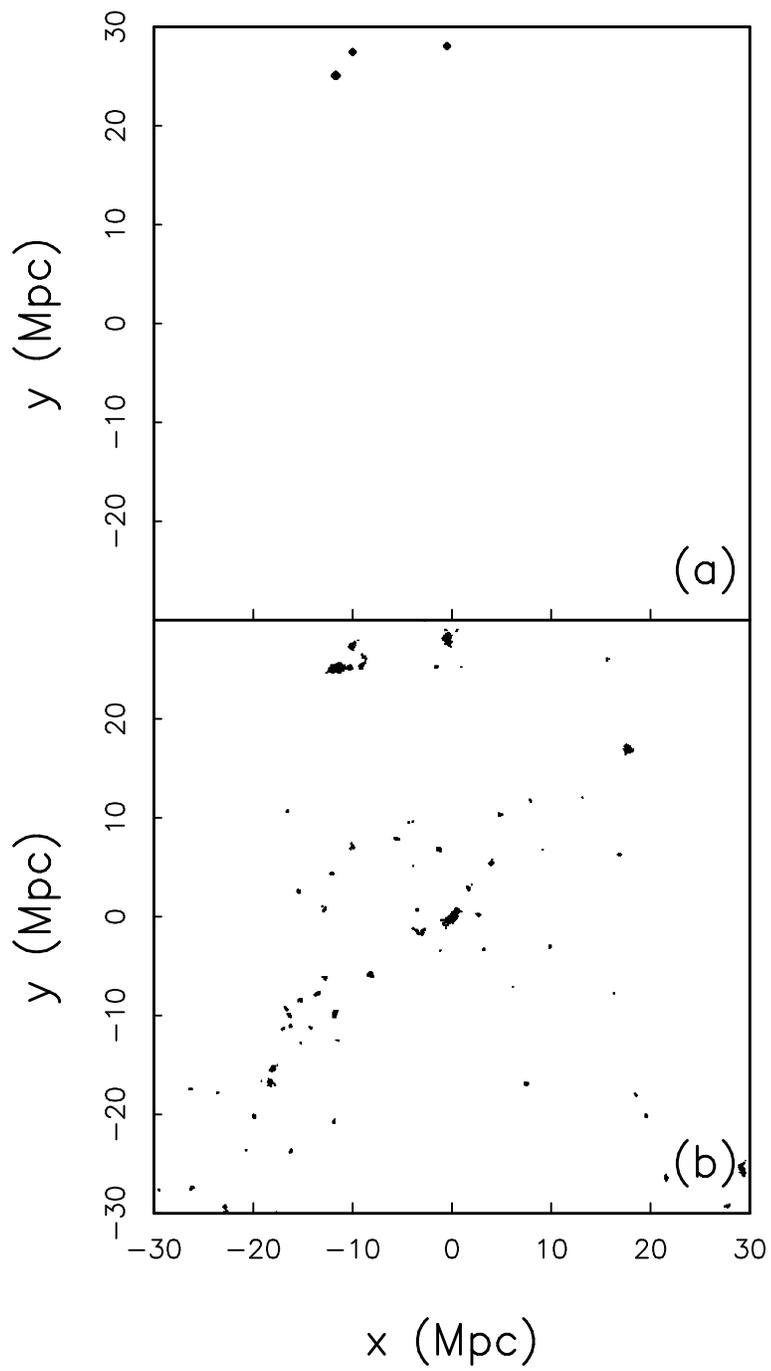
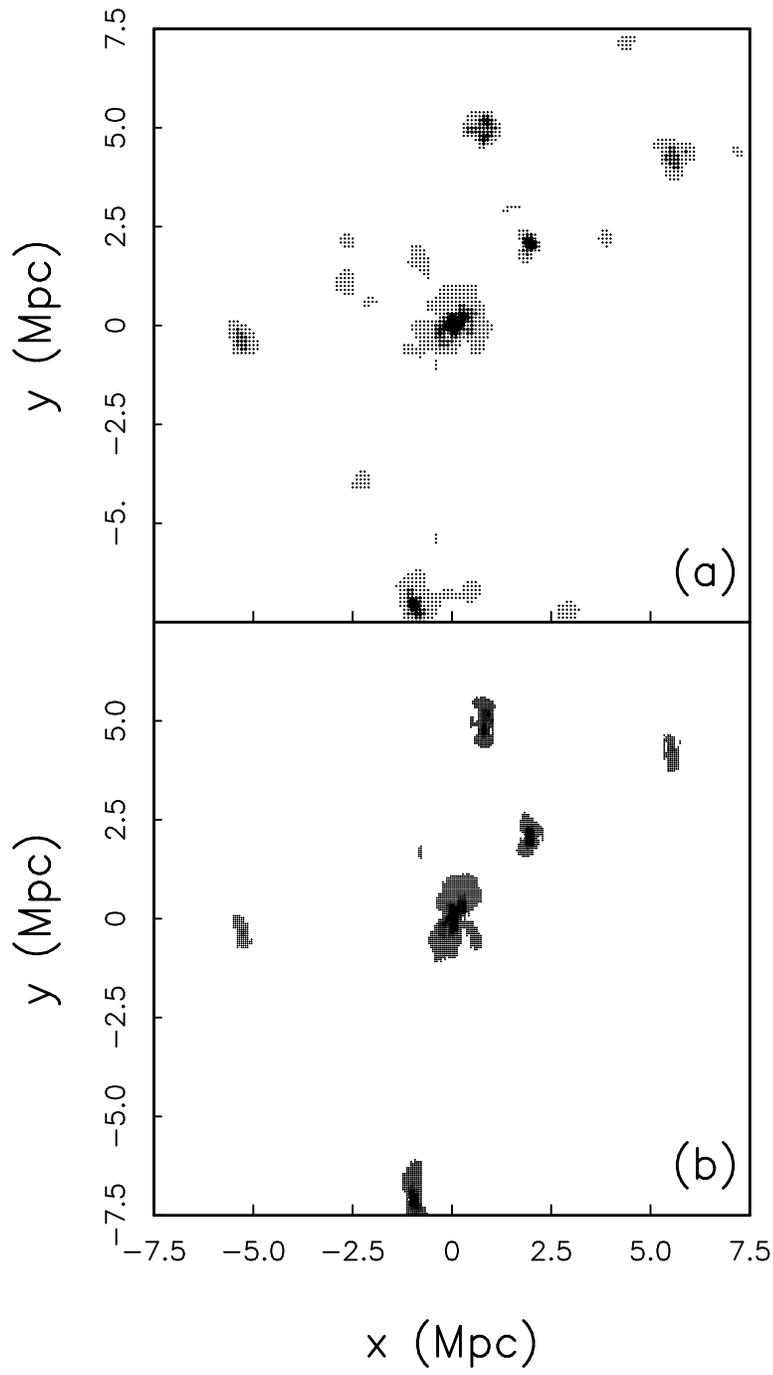
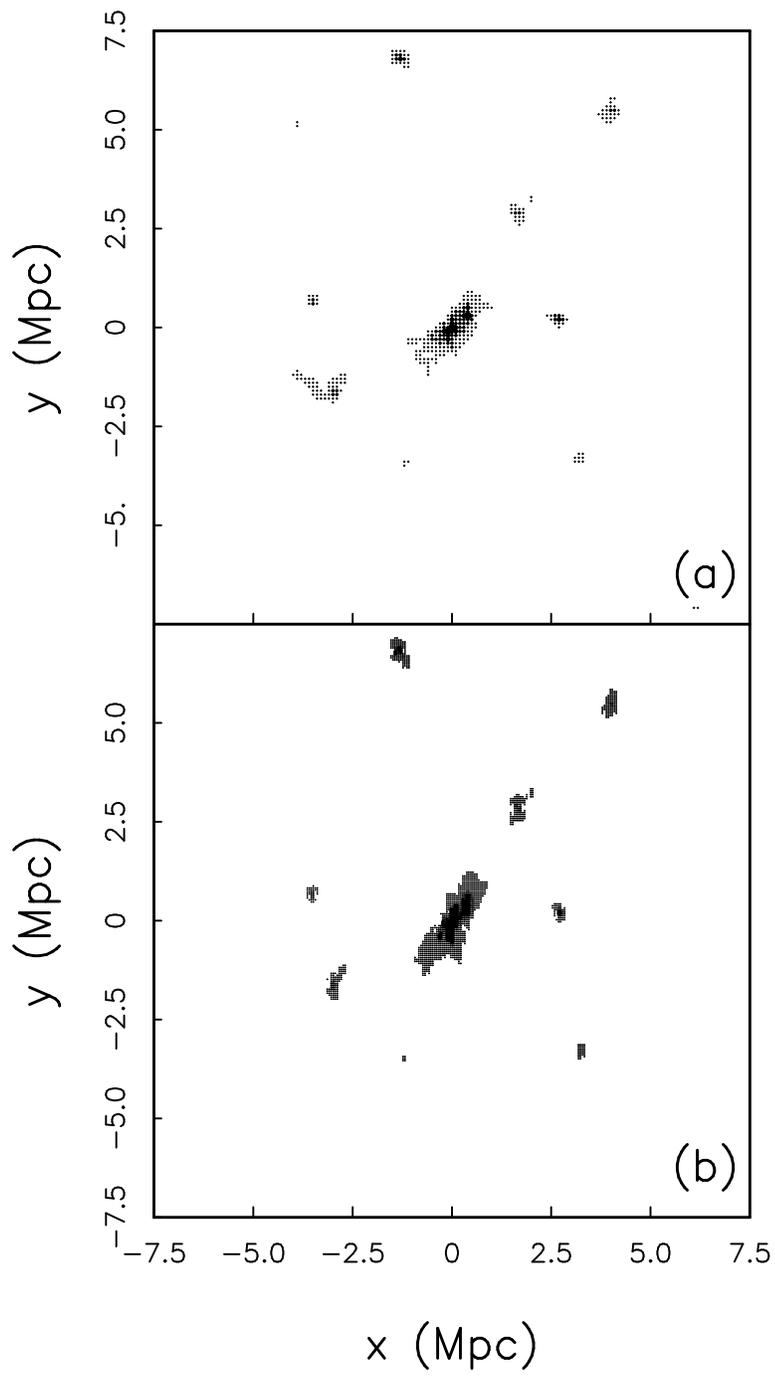
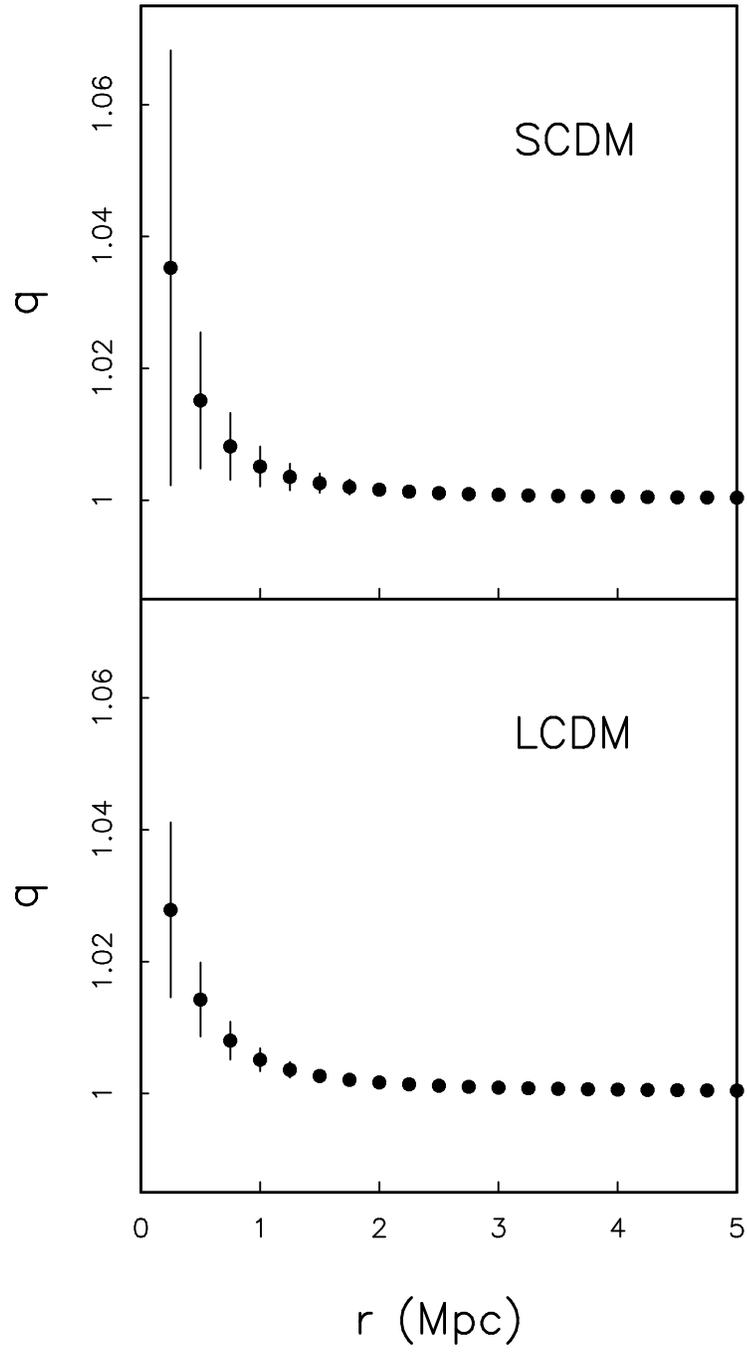
\end{document}